\documentclass[twocolumn,showpacs,preprintnumbers,amsmath,amssymb,superscriptaddress,floatfix]{revtex4}

\usepackage{graphicx,subfigure}
\usepackage{dcolumn}
\usepackage{bm}
\usepackage{psfrag}
\usepackage{multirow}

\def\be{\begin{equation}}       \def\ee{\end{equation}}
\def\bea{\begin{eqnarray}}      \def\eea{\end{eqnarray}}

\begin{document}
\title{Impurity effects in multiferroic compounds}
\author{Trinanjan Datta}
\affiliation{Department of Chemistry and Physics, Augusta State
University, Augusta, GA 30904}
\date{\today}
\begin{abstract}\label{abstract}
We investigate the effect of impurities in multiferroic materials using an equation of motion approach for the spin dynamics of the host multiferroic compound. We model the impurities as a two-level system and focus on the regime where the impurity spins relax slowly. When the impurity strength is weak the host spins oscillate with no decay and the electric polarization is not affected. However as the impurity strength is increased the host spin components get damped and the electrion polarization is suppressed. Since polarization in multiferroic materials is driven by magnetic ordering we conclude that the presence of impurities is detrimental to multiferroicity. 
\end{abstract}

\pacs{75.80.+q, 75.47.Lx, 77.80.-e}
\maketitle
Ferroelectricity in multiferroics is driven by magnetic ordering. The non-collinear spin arrangement below the transition temperature drives the system to a space-inversion and time-reversal broken symmetry state \cite{cheong2007,fiebig}. This causes an electric polarization to be generated. The electric polarization in these materials can be controlled by a magnetic field and the magnetization by an electric field, but are there any factors which are detrimental to this controllability? In this Brief Report, we answer such a question by studying the effect of impurities in multiferroic compounds. Present theoretical \cite{mostovoy,katsura2005,sergienko2006,harris054447,hu2008,fang2009} and computational treatments \cite{sergienko2006} of multiferroic systems have not considered the effect of impurities. 

A multiferroic system with impurities can be described by the microscopic Hamiltonian, $H$, which generates the spin-spiral ground state configuration and includes a coupling of the host spins with the impurity spins\begin{equation}\label{eq:totalhamiltonian}H=H_{mf}+H_{imp}\end{equation}
where $H_{mf}$ is the multiferroic Hamiltonian \cite{katsura2005} and includes the following terms\begin{align}\label{eq:multiferroichamiltonian}
&H_{mf}=H_{ex}+H_{kin}+H_{DM}+H_{an}\\
&H_{ex}=-J\sum_{\langle i,j\rangle}\vec{S}_{i}\cdot\vec{S}_{j}\\
&H_{DM}=-\lambda\sum_{i}(\vec{u}_{i}\times\vec{e}_{z})\cdot(\vec{S}_{i}\times\vec{S}_{i+1})\\
&H_{kin}=\sum_{i}\left(\frac{\kappa}{2}\vec{u}^{2}_{i}+\frac{1}{2M}\vec{P}^{2}_{i} \right)\\
&H_{an}=\sum_{i}D(S^{y}_{i})^{2}
\end{align} In the above Hamiltonian $H_{ex}$ is the Heisenberg exchange Hamiltonian for nearest-neighbor spin interaction, $\vec{S}_{i}$ and $\vec{S}_{i+1}$, of the host multiferroic material. The exchange parameter $J$ between nearest-neighbor spins is considered to be constant. The term $H_{DM}$ represents the Dzyaloshinskii-Moriya interaction and the spin-lattice interaction strength is denoted by $\lambda$. The dynamics of the lattice field is included in $H_{kin}$ where $\vec{u}_{i}$ is the atomic displacement at site $i$ and the canonically conjugate momentum is $\vec{P}_{i}$. In $H_{kin}$, $\kappa$ and $M$ are the spring constant and the effective mass of $u_{i}$ respectively. The easy plane spin-anisotropy term is given by the last expression, $H_{an}$, where the parameter $D>0$. The easy plane in this model is the $x-z$ plane. 

We model the impurities as a two-level system described in terms of an effective spin-1/2 \cite{safanov}. The impurities are assumed to be distributed uniformly throughout the host material and couple with the spins of the multiferroic compound via an exchange interaction. It is assumed that the impurities are diluted enough that they do not form their own fully magnetized sublattice system. We therefore neglect the spin-spin interaction between the impurities. The Hamiltonian, $H^{ex}_{imp}$, of the impurity-multiferroic exchange interaction between the impurity spins and the neighboring host atoms is\begin{align}\label{eq:exchangeimpurity}H^{ex}_{imp}=\sum_{m,\nu}\sum_{\alpha_{m},\alpha}A_{\alpha_{m},\alpha}(\vec{R}_{m},\vec{r}_{\nu})s_{\alpha_{m}}(\vec{R}_{m})S_{\alpha}(\vec{R}_{m}
+\vec{r}_{\nu})\end{align} where the impurity spin components at $\vec{R}_{m}$ is denoted by $\vec{s}(\vec{R}_{m})=(s_{x_{m}}(\vec{R}_{m}),s_{y_{m}}(\vec{R}_{m}),s_{z_{m}}(\vec{R}_{m}))$ and
the spins of the multiferroic system located near the $m$-th impurity is given by
$\vec{S}(\vec{R}_{m}+\vec{r}_{\nu})=(S_{x}(\vec{R}_{m}+\vec{r}_{\nu}),S_{y}(\vec{R}_{m}+\vec{r}_{\nu}),
S_{z}(\vec{R}_{m}+\vec{r}_{\nu}))$. The local co-ordinates for the $m$-th impurity spin are $\alpha_{m}=x_{m},y_{m},z_{m}$ and $\alpha=x,y,z$ are the co-ordinates of the host spin components. The anisotropic exchange parameters of the interaction between $\vec{s}(\vec{R}_{m})$ and $\vec{S}(\vec{R}_{m}+\vec{r}_{\nu})$ are $A_{\alpha_{m},\alpha}(\vec{R}_{m},\vec{r}_{\nu})$ where $\vec{r}_{\nu}$ is the position vector from the
m-th impurity spin to the i-the spin, $\vec{S}_{i}$. 

To obtain a quantum mechanical version of $H^{ex}_{imp}$ we express the impurity spin components in terms of the creation $(c^{\dag}_{m})$ and the annihilation operator $(c_{m})$ respectively. Using canonical commutation relations in Eq.~\ref{eq:exchangeimpurity} we can derive the impurity Hamiltonian, $H_{imp}$, as \begin{align}&H_{imp}=\sum_{m}\Omega_{m}(t=0)n_{m}+ H_{imp}^{ex}\\&H_{imp}^{ex}=\sum_{i,m}\vec{S}_{i}\cdot[\vec{\epsilon}_{z_{m}}(n_{m}-1/2)+ \vec{\epsilon}_{x_{m};y_{m}}\langle c^{\dag}_{m}\rangle\nonumber\\&
+(\vec{\epsilon}^{*}_{x_{m};y_{m}})\langle c_{m}\rangle]\label{eq:impurityhamiltonian}\end{align} where $\Omega_{m}(t=0)$ represents the initial splitting between the energy levels of the two-level system used to model the impurity spins, $n_{m}$ is the population of the impurity spins at that site for a given temperature, $T$. The coupling coefficients are given by the expressions $\epsilon_{x_{m},\alpha;y_{m},\alpha}=\frac{1}{2\gamma}\sum_{\nu}[A_{x_{m},\alpha}(\vec{R}_{m},\vec{r}_{\nu})-iA_{y_{m},\alpha}(\vec{R}_{m},\vec{r}_{\nu})]$ and $\epsilon_{z_{m},\alpha}=\frac{1}{\gamma}\sum_{\nu}A_{z_{m},\alpha}(\vec{R}_{m},\vec{r}_{\nu})$. The coefficient $\epsilon_{x_{m},\alpha;y_{m},\alpha}$ is related to the fast relaxation process and $\epsilon_{z_{m},\alpha}$ is for the slow relaxation process. 

The fast relaxation mechanism accomplishes direct excitation of the impurity ions by the magnetic oscillations of the system. The slow relaxation mechanism is based on the modulation of the impurity energy levels by the oscillations of the magnetic system and is the relaxation process considered in this paper \cite{gurevichmelkov}. At present there is no experimental study in multiferroic compounds which investigates the effects of impurity on spin relaxation rate and its consequences on electric polarization. The consideration of slow relaxation as a mechanism for spin relaxation in multiferroics is therefore an \emph{assumption} in this paper. 

We use Eqs.~(2) - (6) and (7) - (9) to study the effect of impurities on the spin relaxation. For this purpose we implement an equation of motion approach for the spin dynamics \cite{suhl:relaxation,gurevichmelkov} of the system. Consider the equation\begin{align}\label{eq:eom}\frac{d{\vec{S}_{i}}}{dt}= -\gamma{\vec{S}_{i}}\times {\vec{h}_{eff}}\end{align}where $\gamma$ is the gyromagnetic ratio and the effective field, $\vec{h}_{eff}$, is given by\begin{align}&\vec{h}_{eff}=-\frac{\partial H}{\partial \vec{S}_{i}}\nonumber\\ &-\sum_{m}[\vec{\epsilon}_{z_{m}}(n_{m}-1/2)+ \vec{\epsilon}_{x_{m};y_{m}}\langle c^{\dag}_{m}\rangle+(\vec{\epsilon}_{x_{m};y_{m}})^{*}\langle c_{m}\rangle]\end{align} The second term in the above expression arises from the impurity spins. To obtain the effective field contribution we introduce a rotating local co-ordinate system $(\xi,\eta,\zeta)$ and the effective field $\vec{h}_{i}=(h^{\xi}_{i},h^{\eta}_{i},h^{\zeta}_{i})$ in this frame of reference \cite{nagamiya}. We first obtain the contribution from the $H_{mf}$ part of the Hamiltonian and find \begin{align} &h^{\xi}_{i}=4JS^{\xi}_{i}
\cos qa\nonumber\\&+\frac{\lambda^{2}(S^{\zeta}_{i})^{2}}{\kappa}
\sin^{2}qa(S^{\xi}_{i+1}+S^{\xi}_{i-1})\\
&h^{\eta}_{i}=2JS^{\eta}_{i}-2DS^{\eta}_{i}\\
&h^{\zeta}_{i}=2S^{\zeta}_{i}\left(J(q)+\frac{\lambda^{2}(S^{\zeta}_{i})^{2}}{\kappa}\sin^{2}qa\right)
\end{align} where $q$ is the wavector and $a$ is the lattice spacing. 

The contribution to the effective field from the impurity part of the Hamiltonian can be obtained by considering the kinetics of the $m$-th impurity spin population  
\begin{align}
\frac{d}{dt}n_{m}=-\Gamma_{\parallel,m}(\Omega_{m})[n_{m}-n_{T}(\Omega_{m})]
\end{align}
where $\Gamma_{\parallel,m}(\Omega_{m})$ is a relaxation coefficient for the impurity spins. At each instant of time the population $n_{m}=\langle c^{\dag}_{m}c_{m}\rangle$ relaxes to the equilibrium value $n_{T}(\Omega_{m})=[\exp(\Omega_m / k_{B}T)+1]^{-1}$ corresponding
to the dynamic splitting $\Omega_{m}=\Omega_{m}(t=0)+\delta\Omega^{i}_{m}(t)$ where $\delta\Omega^{i}_{m}(t)\equiv \vec{\epsilon}_{z_{m}}\cdot\vec{S}_{i}$. For the case of \emph{slow relaxation} the impurity spins modulate slowly,
$|\delta\Omega^{i}_{m}|/k_{B}T<<1$, and the impurity contribution to the effective magnetic field is \cite{safanov} \begin{align}\label{eq:impurityeffectivefield}&\delta\vec{h}_{eff}=-\sum_{m}\vec{\epsilon}_{z_{m}}\left[n_{T}(\Omega_{m}(0))-\frac{1}{2}\right]\nonumber\\&-\sum_{m}\vec{\epsilon}_{z_{m}}\frac{\partial n_{T}(\Omega_{m}(0))}{\partial\Omega_{m}(0)}\left[\vec{\epsilon}_{z_{m}}\cdot\vec{S}_{i}(t)\right]\nonumber\\
&+\sum_{m}\vec{\epsilon}_{z_{m}}\frac{\partial n_{T}(\Omega_{m}(0))}{\partial\Omega_{m}(0)}\frac{1}{\Gamma_{\parallel,m}\Omega_{m}(0)}\left[\vec{\epsilon}_{z_{m}}\cdot\frac{d\vec{S}_{i}(t)}{dt}\right]\end{align}The third term in the above equation, Eq.~\ref{eq:impurityeffectivefield}, is responsible for generating relaxation and the first and the second term introduces an anistropy. We will neglect the effect of the anisotropies generated by these terms and focus on the effect of the relaxation term. 

\begin{figure}[t]
\centering 
 \subfigure[$\beta=0.1$]{\label{subfig:relax1}
   \includegraphics[width=2.5in]{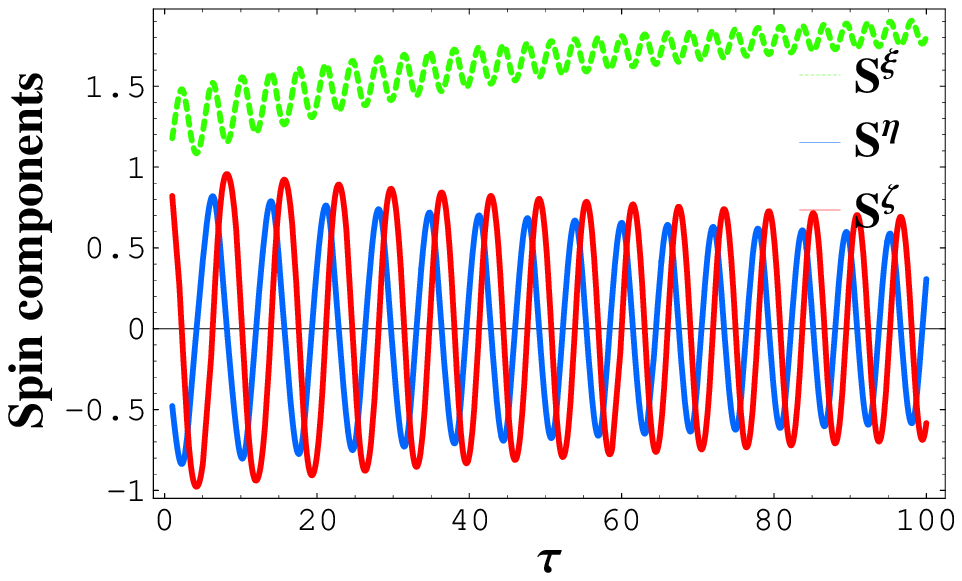}}
   \subfigure[$\beta= 1$]{
   \includegraphics[width=2.5in]{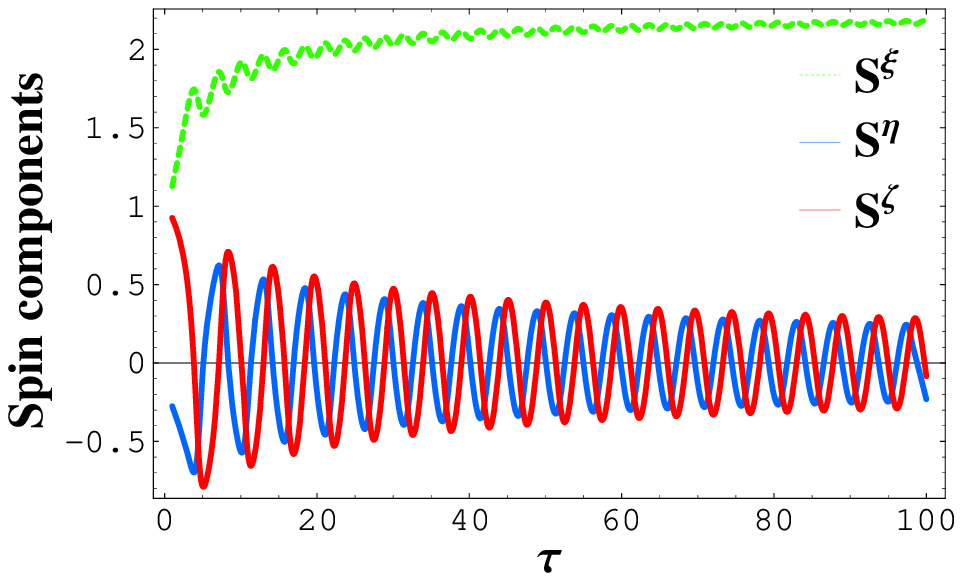}}
  \subfigure[$\beta= 10$]{\label{subfig:alasbandstruct}
  \includegraphics[width=2.5in]{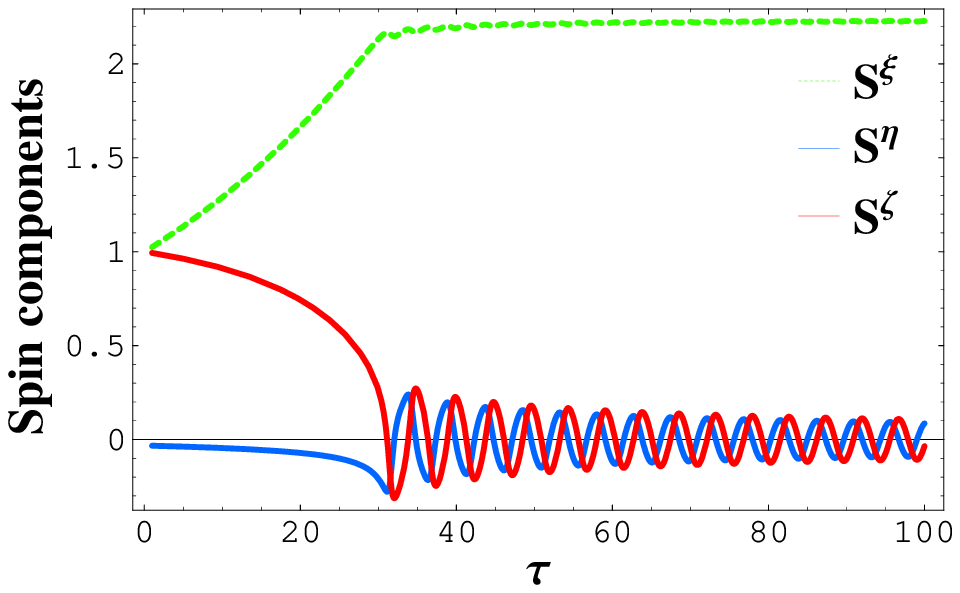}}
\caption{(Color online) Relaxation of spin components, $(S^\xi,S^\eta,S^\zeta)$, for various values of the damping coefficient $\beta$. The range for $\beta$ is varied from $0.1$ to $10$ by a factor of ten. The spin component, $S^{\zeta}$, responsible for generating the electric polarization (see Eq.~\ref{eq:polarization}) along the $x-$axis is damped leading to a suppression of the induced ferroelectricity.}
\label{fig:1}
\end{figure}

Using Eq.~\ref{eq:impurityeffectivefield} and the equation of motion (Eq.~\ref{eq:eom}) contribution for the spin components in the rotating frame, $(S^{\xi}_{i},S^{\eta}_{i},S^{\zeta}_{i})$, the effective magnetic field can be written as\begin{align}
&\frac{d S^{\xi}_{i}}{d\tau}=-2DS^{\eta}_{i}S^{\zeta}_{i}\label{eq:equationofmotion1}\\
&\frac{d S^{\eta}_{i}}{d\tau}=-4J S^{\zeta}_{i}S^{\xi}_{j}\cos^{2}\frac{qa}{2}+ \gamma\beta S^{\zeta}_{i}\frac{d S^{\xi}_{i}}{d\tau}\label{eq:equationofmotion2}\\
&\frac{d S^{\zeta}_{i}}{d\tau}=2\left(D+2J\cos^{2}\frac{qa}{2}\right)S^{\xi}_{i}S^{\eta}_{i}-\gamma\beta S^{\eta}_{i}\frac{d S^{\xi}_{i}}{d\tau}
\label{eq:equationofmotion3}\end{align}
where $\tau=\beta t$. The damping co-efficient $\beta=\left( \frac{\epsilon^{2}c_{imp}}{\Gamma_{||}k_{B}T}\right)\frac{\exp(\Omega_o / k_{B}T)}{[\exp(\Omega_o / k_{B}T)+1]^2}$ where $k_{B}$ is the Boltzmann constant. The impurity concentration, $c_{imp}=\gamma n_{imp}z_{imp}$ where $n_{imp}$ is the number of impurities per unit volume and $z_{imp}$ is the average number of magnetic neighbors for one impurity. The damping coefficient is dependent on the impuritiy concentration $(c_{imp})$, the temperature $(T)$, the energy scale of the impurities $(\Omega_{m}(t=0))$ taken to be a constant value $\Omega_{m}(t=0)=\Omega_{o}$, and the impurity coupling coefficient $\vec{\epsilon}_{m}=(\epsilon,0,0)$.

In deriving the spin dynamics of the multiferroic system we have considered the following facts. The original effective magnetic field generates terms which couple the $u_{y}$ displacement field. However, in the polarized state there is no polarization along the $b$-axis, so the expectation value of the displacement field along that direction is zero, $\langle u_{y}\rangle=0$. We therefore ignore the contribution of these terms in the derivation of the equation of motion. In the equation of motions, Eqs.~\ref{eq:equationofmotion1}-~\ref{eq:equationofmotion3}, we have ignored terms greater than second order in the spin components. Also, the effective magnetic field terms which are proportional to the $\lambda$ coupling do not appear in the final expressions as these terms belong to a fourth order spin contribution and are neglected. Therefore, in the present theory (second order in the spin components) the spin-lattice interaction has no role to play. 

We solve Eqs.~\ref{eq:equationofmotion1}-~\ref{eq:equationofmotion3} for appropriate initial conditions to obtain Fig~\ref{fig:1}. To begin with we consider the spiral in the $x-z$ plane. Parameters relevant for the multiferroic material TbMnO$_3$ \cite{kajimoto} are used for the simulation: $J$ = 0.15 meV, and $D$ = 0.4 meV. We choose $qa=\pi/4$ and $\gamma=2$. From Fig.~\ref{fig:1} we observe that as the contribution of the $\beta-$parameter is increased their is a pronounced damping of the $S^{\eta}_{i}$ and $S^{\zeta}$ spin components (see Fig.~\ref{fig:1}). In the presence of damping the $S^{\xi}$ component reaches an equilibrium value and its oscillations are suppressed also.  

Electric polarization, $\vec{P}$, in multiferroic materials can be generated by a spin current,
$\vec{j}_{s}\propto \vec{S}_{i}\times\vec{S}_{j}$ \cite{katsura2005}. The polarization vector defined as $\vec{P}\propto \vec{e}_{ij}\times(\vec{S}_{i}\times\vec{S}_{j})$, implies that the spin current for the spin arrangement chosen in this calculation should be along the $y-$axis and the polarization along the $x-$axis. Using the expression for spin components in the rotating frame \cite{nagamiya} we have for the magnitude of the $x-$component of $\vec{P}$\begin{equation}\label{eq:polarization} |P_{x}|=[S^{2}+(S^{\zeta})^{2}]\sin(\Delta\theta)\end{equation}where
we have used the fact that $S^{\zeta}_{i+1}\approx S^{\zeta}_{i}$ and set either component to $S^{\zeta}$, $\Delta\theta=\theta_{i}-\theta_{i+1}$ is the angular difference between the spins in the nearest neighbor sites. In the above equation we have replaced the $S^{\xi}$ component of the spin with an equilibrium value, S, as seen in Fig.~\ref{fig:1}. From Eq.~\ref{eq:polarization} we observe that as the damping coefficient is increased, a suppression of the $S^{\zeta}$ component (see Fig.~\ref{fig:1}) occurs leading to a suppression of the electric polarization, $|P_{x}|$. 

Based on the above analysis we conclude that impurities are detrimental for multiferroic materials. Impurities lead to a damping of the spin polarization compent amplitudes and in turn to a suppression of ferroelectric polarization. 

TD acknowledges helpful discussions with Kingshuk Majumdar. 
\bibliography{spinrelaxation}
\end{document}